\documentclass{article}

\usepackage[preprint]{neurips_2025}

\usepackage[utf8]{inputenc} % allow utf-8 input
\usepackage[T1]{fontenc}    % use 8-bit T1 fonts
\usepackage{hyperref}       % hyperlinks
\usepackage{url}            % simple URL typesetting
\usepackage{booktabs}       % professional-quality tables
\usepackage{amsfonts}       % blackboard math symbols
\usepackage{nicefrac}       % compact symbols for 1/2, etc.
\usepackage{microtype}      % microtypography
\usepackage{xcolor}         % colors

\usepackage{orcidlink}

\title{Ergonomic Assessment of Work Activities for an Industrial-oriented Wrist Exoskeleton}

\author{%
  Roberto F. Pitzalis \orcidlink{0000-0003-1464-6219} \\
  XoLab, Advanced Robotics (ADVR)\\
  Istituto Italiano di Tecnologia (IIT)\\
  Genoa, Italy \\
  \\
  DIME \\
  University of Genoa\\
  Genoa, Italy \\
  \texttt{roberto.pitzalis@iit.it} \\
  \And
  Nicholas Cartocci \orcidlink{0000-0002-4908-9769} \\
  XoLab, Advanced Robotics (ADVR)\\
  Istituto Italiano di Tecnologia (IIT)\\
  Genoa, Italy \\
  \\
  DIBRIS \\
  University of Genoa\\
  Genoa, Italy \\
  \texttt{nicholas.cartocci@iit.it} \\
  \And
  Christian Di Natali \orcidlink{0000-0001-7399-7399} \\
  XoLab, Advanced Robotics (ADVR)\\
  Istituto Italiano di Tecnologia (IIT)\\
  Genoa, Italy \\
  \texttt{christian.dinatali@iit.it} \\
  \And
  Luigi Monica \orcidlink{0000-0001-5702-6912} \\
  Technological Innovation and Safety Equipment\\
  INAIL\\
  Rome, Italy \\
  \texttt{l.monica@inail.it} \\
  \And
  Darwin G. Caldwell \orcidlink{0000-0002-6233-9961} \\
  Advanced Robotics (ADVR)\\
  Istituto Italiano di Tecnologia (IIT)\\
  Genoa, Italy \\
  \texttt{darwin.caldwell@iit.it} \\
  \And
  Giovanni Berselli \orcidlink{0000-0003-0093-3006} \\
  DIME \\
  University of Genoa\\
  Genoa, Italy \\
  \texttt{giovanni.berselli@unige.it} \\
  \And
  Jesús Ortiz \orcidlink{0000-0001-9475-1945} \\
  XoLab, Advanced Robotics (ADVR)\\
  Istituto Italiano di Tecnologia (IIT)\\
  Genoa, Italy \\
  \texttt{jesus.ortiz@iit.it} \\
}

\begin{document}

\maketitle

\begin{abstract}
Musculoskeletal disorders (MSD) are the most common cause of work-related injuries and lost production involving approximately 1.7 billion people worldwide and mainly affect low back (more than 50\%) and upper limbs (more than 40\%). It has a profound effect on both the workers affected and the company. This paper provides an ergonomic assessment of different work activities in a horse saddle-making company, involving 5 workers. This aim guides the design of a wrist exoskeleton to reduce the risk of musculoskeletal diseases wherever it is impossible to automate the production process. This evaluation is done either through subjective and objective measurement, respectively using questionnaires and by measurement of muscle activation with sEMG sensors.
\end{abstract}

\section{Introduction}

Prolonged, violent, irregular motions and incorrect postures in the working environment can cause a variety of common musculoskeletal diseases, which can be episodic or chronic in duration. They can progress from mild to severe and impair the quality of life of workers, by reducing mobility and dexterity in activities of daily living (ADL). As reported by the \citet{euosha2020working} the majority of these accidents affects low back (more than 50\%) and upper limbs (more than 40\%). In this end, exoskeletons and wearable robotics technologies are gaining attention to support workers in strenuous manual activities wherever it is impossible to automate the production process. However, an ergonomic assessment of the working environment and activity is necessary to design these devices effectively, guaranteeing reliability, compliance, accurate assistance, and safety, and justify the introduction of an assistive device. \\
This study aims to provide an ergonomic assessment of the musculoskeletal risk to the hand and wrist, arising from work activities in a horse saddle production plant. Since most of the activities are handcrafts and require excellent manual skills, which cannot be transferred to a robot, exoskeletons specifically designed to reduce biomechanical overload on workers may form an excellent technological alternative (\citet{Armstrong1987,Crea_2021,Monica_2021}).

\section{Methodology}

\begin{figure}[htb!]
    \centering
    \includegraphics[width=\textwidth]{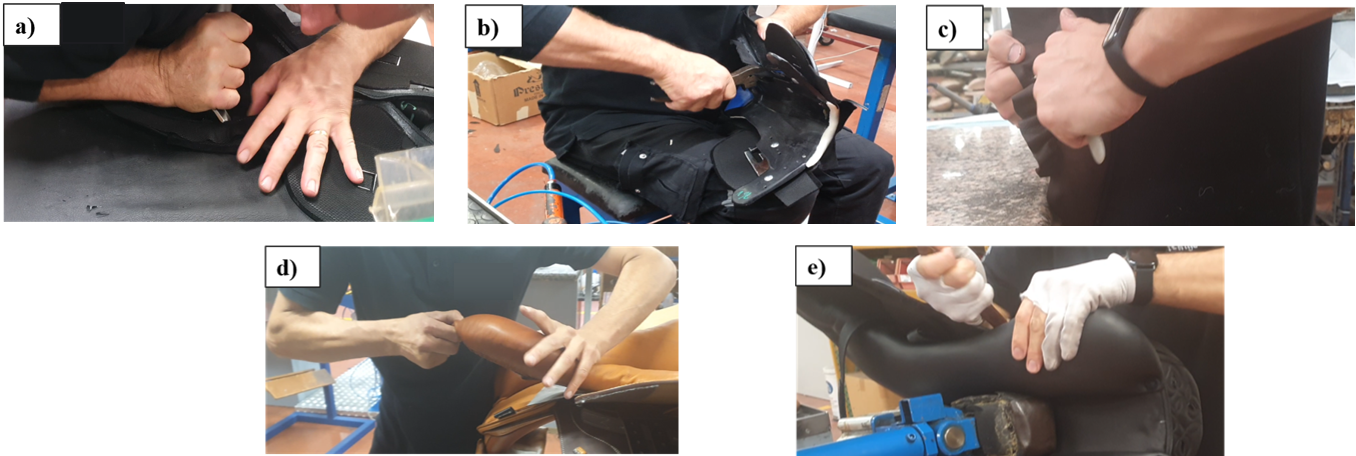}
    \caption{Working tasks assessed: seat preparation (a) and assembly (b), cushion covering (c), cushion mounting (d), and sewing (e).}
    \label{fig:Working_tasks}
\end{figure}

In this study, the participants are craftsmen skilled in the production of horse saddles. The activities done in their manufacturing plant are laborious, and one of the major causes of injuries (e.g. carpal tunnel syndrome) and absence from work. Five healthy male subjects (age range: [25, 45] years) were recruited in the same day. Four of them are identified as right-handed and only one as left-handed (\citet{pitzalis2024development}). The experiment was conducted in accordance with the Declaration of Helsinki and was approved by the Ethics Committee of Liguria (protocol reference number: CER Liguria 001/2019). The protocol respects users’ right to privacy in accordance with the General Data Protection Regulation (GDPR), \citet{goddard2017eu}. \\
Five working tasks, from ’a’ to ’e’ as shown in Figure \ref{fig:Working_tasks}, are identified by the industrial hygienist of the company to be more strenuous during the work shift. Not all the subjects performed the same activities, in particular: the participants P1 and P3 conducted tasks ’a’ and ’b’; the participants P2, P4, and P5 conducted tasks ’c’, ’d’, and ’e’, respectively. All these activities are different but they have in common the use of force with the hand and wrist, the placement of the wrist in non-ergonomic positions, and the grasping of a work tool.

Before each experimental session, Maximum muscle Voluntary Contraction (MVC) is measured for the following 5 wrist activities: wrist flexion, wrist extension, wrist radial deviation, wrist ulnar deviation, and grasping. This provides a baseline to be compared with the muscle activation during actual work. To ensure correct parallelism with the ground, each participant is seated comfortably on a chair and with the elbow and forearm placed on a table, leaving only the wrist free to move (\citet{sanchoelectromyography}). Participants receive visual instructions asking them to apply as much force as possible against a contrasting element while performing a wrist gesture. To detect and measure the hand/wrist muscular effort made during the work, the electrical muscle activation of each subject’s forearm is recorded with a Myo Armband (Thalmic Labs), which is a commercially available 8-channel, dry-electrode, low-sampling rate (200 Hz), and low-cost electromyography (EMG) sensor. Because electromyography signals are notoriously noisy, they are pre-treated before being used (\citet{pitzalis2024development}). They are first rectified, then filtered with a cutoff frequency of 5 Hz, and finally normalized to the 98\% percentile of the MVC to make them comparable among different subjects. \\
For each experimental session, participants perform their typical work activity for 3 to 5 minutes, while muscle activation measurements are acquired using EMG sensors on both the left and right arms. Once the task is completed, under the administrator’s supervision, participants are asked to complete a series of scales and questionnaires for subjective validation of the perceived level of fatigue experienced and effort required while performing the task. These assessments included: NASA Task Load Index (TLX), the American Conference of Governmental Industrial Hygienists (ACGIH) Threshold Limit Value (TLV) for Hand Activity (HAL), and Visual Hand Pain Mapping (VHPM).

\section{Results}
\label{Results}

\begin{figure}[ht!]
    \centering
    \includegraphics[width=\textwidth]{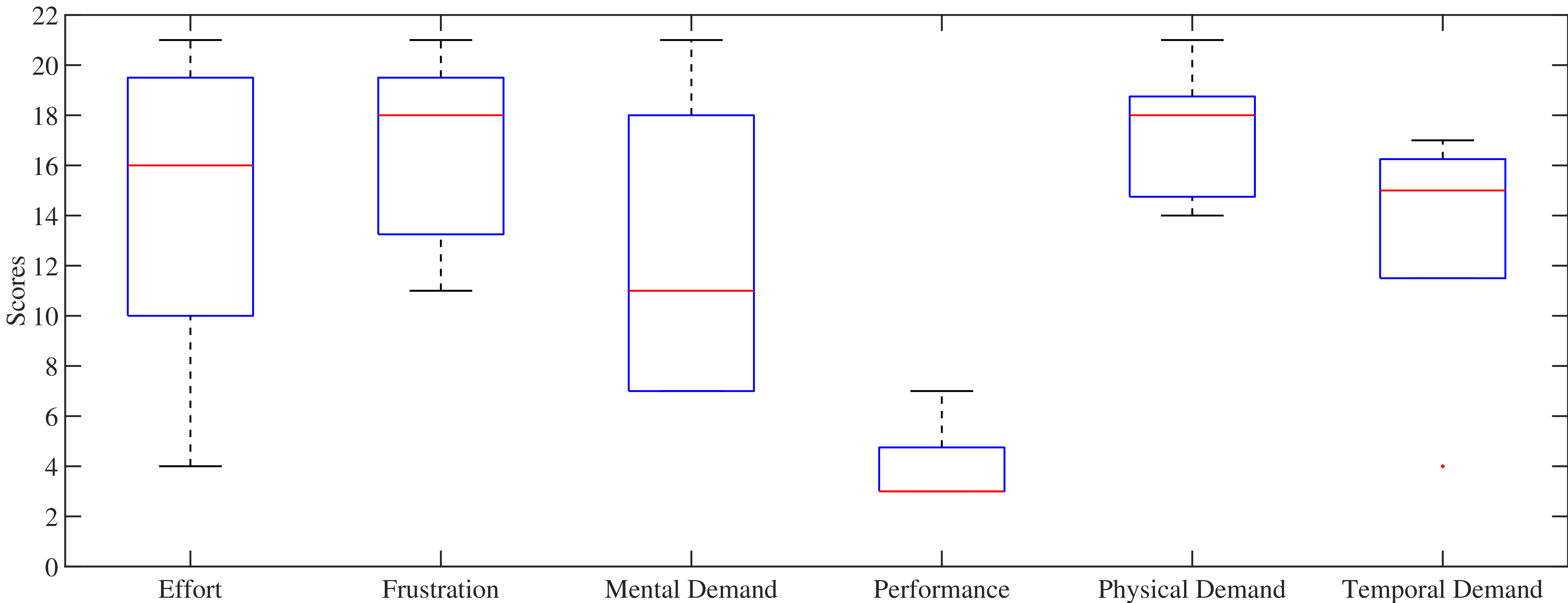}
    \caption{NASA Task Load Index is divided into six subjective sub-scales (Mental Demand, Physical Demand, Temporal Demand, Performance, Effort, Frustration). These scales score perceived fatigue from "Very Low"(1) to "Very High" (21), except for "Performance", with which a task is performed, which is rated from "Perfect" (1) to "Failure" (21). In all cases lower values represents better scores.}
    \label{fig:NASA_TLX}
\end{figure}

Data from the questionnaires (subjective assessment of perceived fatigue) and the measurements of electrical muscle activation (objective assessment of muscle fatigue) are collected and combined to search for a correlation between what is perceived by the subject and what is measured. This analysis gives a more objective perspective on the impact of that task on the worker’s health. It is an essential step in preventing occupational accidents. \\
NASA TLX is a widely used subjective assessment tool that rates perceived workload (\citet{hart1988development}). The data acquired by administering the NASA TLX are reported in Figure \ref{fig:NASA_TLX}. It can be seen that subjects tend to define the tasks as moderately or highly strenuous, performing them at a near-optimal level of performance. 

\begin{figure}[ht!]
    \centering
    \includegraphics[width=\textwidth]{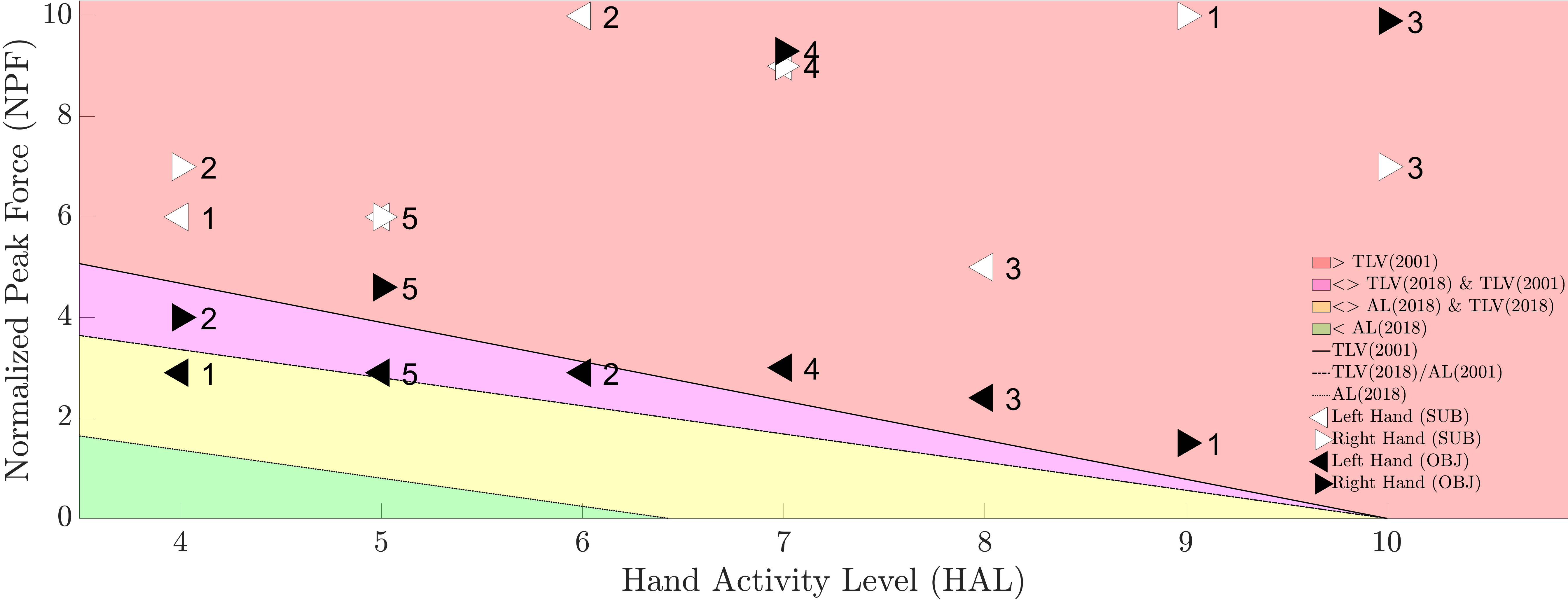}
    \caption{HAL/ACGIH TLV assessment. The graph shows the maximum normalized force values (NPF) of 5 subjects exerted by the right hand and left hand while performing work activities, reporting both the old (2001) and updated (2018) threshold limits. White triangles report subjective (SUB) measures assessed by administering a CR-10 Borg scale to subjects; while black triangles report objective (OBJ) measures derived from an analysis of EMG data.}
    \label{fig:HAL_NPF}
\end{figure}

As suggested in the technical report ISO/TR 12295:2014 for the application of ISO ergonomic standards for manual handling of loads (ISO 11228) and static working postures (ISO 11226), \citet{iso11228_1,iso11228_3}, the HAL/ACGIH TLV may provide insight on the assessment of musculoskeletal risk factors associated with the hand and wrist during multitask work. It estimates the level of manual activity in terms of repetitiveness of movements and relative duration of the task compared to breaks, and the intensity of the applied force, which can be assessed with EMG or, more simply, with the CR-10 (Borg) scale. To compare between objective and subjective perception, the applied force is evaluated using respectively the EMG measurements and the Borg scale. Hereafter, the results of the HAL/ACGIH TLV analysis are presented in Figure \ref{fig:HAL_NPF}, reporting both the old threshold limits (dated back to 2001) and the updated version (2018), \citet{yung2019modeling}. As can be seen, the level of action limits (AL) has been lowered in the new version, thus widening the risk zone (yellow + purple in Figure \ref{fig:HAL_NPF}). All activities rated below the AL line are considered non-hazardous; those rated between the AL and TLV limits are of significant risk and require monitoring; while those above the TLV (red area in Figure \ref{fig:HAL_NPF}) are at high risk and require timely action to improve working conditions. \\
Overall subjectively perceived fatigue by workers is high and above the threshold limit value (white triangles in Figure \ref{fig:HAL_NPF}). Starting from the EMG measurements, the Normalized Peak Force (NPF) has been estimated considering the average value of the predominant channel during the task normalized for its MVC. Objectively, there is a tendency to detect medium to high fatigue, with values around the threshold limit, but still lower than what the person perceives at the subjective level.

Finally, following the guidelines proposed by the VHPM protocol (\citet{moellhoff2023visualization}), subjects were asked to indicate and judge in which areas of their hands and wrists they experienced the major discomfort during the activity.

\section{Discussion}

The results obtained correlate worker-perceived fatigue and electromyography sensor measurements. As known in the literature, amplitude-based measures such as the maximum value, the average rectified value and the root mean square value of the EMG signal have generally been shown to increase with fatigue, leading to a decline in the muscle output force and performance exerted by a limb. Despite the limitations of this work, such as short duration of the task-performing, the number of participants and repetitions, which will be extended in future, the results of the HAL/ACGIH TLV assessment reported in Section \ref{Results} show that subjects tend to perceive greater fatigue than is actually objectively detected at the level of muscle activity. This difference can also be explained by the fact that the NASA TLX results show that subjects, in addition to physical exertion, also perceive medium to high mental demand, temporal demand, and
frustration, which cannot be measured by processing muscle activation signals. Overall, the muscle fatigue to which workers are subjected is medium to high, and through a VHPM analysis, they highlight discomfort in their hands and wrists at critical points such as the carpal bones of the wrist, ulna and radius joints, in the thumb, the index and middle fingers. These are considered significant sites of interest for the onset of musculoskeletal disorders such as carpal tunnel syndrome, strains, tendinitis, and tenosynovitis. These findings are of enormous value in designing a wrist exoskeleton, selecting movements to actuate, choosing sensors, and positioning them according to the most active forearm muscles, as well as for the design of exoskeleton garments with the addition of elastic, soft or hard interfaces to protect the most delicate parts of the limb.

\section{Conclusions}

This study forms the basis for an ergonomic assessment of working tasks involving hand/wrist manipulation actions, either through subjective (questionnaires, scales and personal opinions) or objective measurements (sEMG data). Results can be effective in identifying critical tasks for worker’s health and may suggest that adopting new technological solutions could prevent or reduce the onset of work-related musculoskeletal disorders, although the application of exoskeletons to task involving precise hand movements is still a challenging area. To this end, future work include the design of an exoskeleton to assist the most frequent wrist movements performed by workers and the addition of interfaces to protect the most delicate parts of the limb.

\begin{ack}
Our thanks go out to INAIL (Italian Workers Compensation Authority) for its collaboration and funding in the development of novel types of industrial exoskeletons, and to the company Prestige Italia S.p.A., where the data were collected, for its willingness to improve workers’ health conditions.
\end{ack}

%%%%%%%%%%%%%%%%%%%%%%%%%%%%%%%%%%%%%%%%%%%%%%%%%%%%%%%%%%%%

\bibliographystyle{plainnat}
\bibliography{bibl}

%%%%%%%%%%%%%%%%%%%%%%%%%%%%%%%%%%%%%%%%%%%%%%%%%%%%%%%%%%%%

% \appendix

% \section{Technical Appendices and Supplementary Material}
% Technical appendices with additional results, figures, graphs and proofs may be submitted with the paper submission before the full submission deadline (see above), or as a separate PDF in the ZIP file below before the supplementary material deadline. There is no page limit for the technical appendices.

%%%%%%%%%%%%%%%%%%%%%%%%%%%%%%%%%%%%%%%%%%%%%%%%%%%%%%%%%%%%

\end{document}